\documentstyle[aps,twocolumn,prl,tighten,flushrt]{revtex} %eqsecnum,
\input epsf
\def\plotone#1{\centering \leavevmode
\epsfxsize= 0.7\columnwidth \epsfbox{#1}}

\def\be{\begin{equation}}
\def\ee{\end{equation}}
\def\bea{\begin{eqnarray}}
\def\eea{\end{eqnarray}}

\def\muk{\mu{\rm K}}

\def\cmm2{{\,\rm cm^{-2}}}
\def\cm2{{\,{\rm cm}^2}}
\def\cmm3{{\,{\rm cm}^{-3}}}
\def\gcmm3{{\,{\rm g\,cm^{-3}}}}

\def\fun#1#2{\lower3.6pt\vbox{\baselineskip0pt\lineskip.9pt
  \ialign{$\mathsurround=0pt#1\hfil##\hfil$\crcr#2\crcr\sim\crcr}}}
\def\C{{\cal C}}

\def\'{^{\prime}}

\def\c{{\cal C}}

\def\p3m{P$^3$M}

\def\la{\mathrel{\mathpalette\fun <}}

\def\fun#1#2{\lower3.6pt\vbox{\baselineskip0pt\lineskip.9pt
  \ialign{$\mathsurround=0pt#1\hfil##\hfil$\crcr#2\crcr\sim\crcr}}}
\begin{document}
\twocolumn[\hsize\textwidth\columnwidth\hsize\csname @twocolumnfalse\endcsname
\preprint{}
\title{Cosmic Microwave Background Anisotropy Window Functions Revisited}
\author{Lloyd Knox}
\date{\today}
\maketitle

\begin{abstract}
  The primary results of most observations of cosmic microwave
  background (CMB) anisotropy are estimates of the angular power
  spectrum averaged through some broad band, called band-powers.
  These estimates are in turn what are used to produce constraints on
  cosmological parameters due to all CMB observations.  Essential to
  this estimation of cosmological parameters is the calculation of the
  expected band-power for a given experiment, given a theoretical
  power spectrum.  Here we derive the "band-power" window function
  which should be used for this calculation, and point out that it is
  not equivalent to the window function used to calculate the
  variance.  This important distinction has been absent from much of
  the literature: the variance window function is often used as the
  band-power window function.  We discuss the validity of this assumed
  equivalence, the role of window functions for experiments that
  constrain the power in {\it multiple} bands, and summarize a
  prescription for reporting experimental results.  The analysis
  methods detailed here are applied in a companion paper to three
  years of data from the Medium Scale Anisotropy Measurement.
\end{abstract}
%]
\pacs{Valid PACS appear here.}
\narrowtext
]

\section{Introduction}

Measurement of the anisotropy of the Cosmic Microwave Background
(CMB) is proving to be a powerful cosmological probe.  However, exact
statistical treatment of the data is
complicated and time-consuming, and promises to become prohibitively
so in the very near future.  This difficulty explains why
no one has calculated the likelihood of cosmological parameters,
given the available data from all CMB experiments.  

Instead, constraints on cosmological parameters have been derived by
approximate methods---namely, the use of ``radical compression''
\cite{BJKII}.  Reduction of CMB data to estimates of the angular power
spectrum, $C_l$, can be viewed as a form of data compression.  This
compressed data is then in turn used to constrain cosmological
parameters.  Usually, the compression is not to estimates of the
individual $C_l$s themselves, but to band-powers \cite{bondbp} which
are averages of the power spectrum through a certain filter, or window
function.  

These band-powers, together with their window functions, have
traditionally been the main results of CMB experiments.
Unfortunately, a large number of experimental results papers only
provide what we will call the variance window function, and not the
``band-power'' window function.  Indeed, the distinction between the
two has been missing in much of the literature.  They are not
equivalent, except in the limit of vanishing off-diagonal signal
correlations.  Reports of constraints on the CMB power spectrum should
contain the latter type of window function, together with the
quantification of the uncertainty in the band-power.

Not all reductions of CMB data to power spectrum estimates have been
presented with only variance window functions.  As a rough guide, those
that have been analyzed with "quadratic estimators" have the right
(band-power) window function while those that have been analyzed by
evaluation of the likelihood function do not.  The persistence of this
confusion is probably due to the fact that
likelihood analysis obscures the relation between the data and the
derived band-power, which is much clearer when one uses an estimator.
It has been shown \cite{BJKI} that a particular quadratic estimator
\cite{BJKI,teg,hamiltona} is guaranteed to produce the
maximum-likelihood result (if used iteratively) and below we exploit
this fact to derive the expression for the band-power window function
appropriate for likelihood analysis.

In a companion paper \cite{companion} we apply the radical compression
procedures detailed here to three years of data from the Medium Scale
Anisotropy Measurement (MSAM).  This procedure is a combination of techniques
developed in \cite{BJKI}, \cite{BJKII} and here.  The application to
MSAM strongly demonstrates the power of this method compared to the
usual approach of compression to flat band-powers.  In particular, the
analysis in the companion paper results in a new and significant
constraint at $l \sim 400$, a theoretically very interesting region of
the power spectrum, which had been previously obscured by use of 
variance window functions, rather than band-power window functions.

\section{The Band-power Method}

A very useful meeting point for theory and experiment is provided
by the band-power.  An important property of a meeting point,
is that both parties planning on meeting should be able to
get there.  Although the directions for going from the data
to the band-power are already clearly explained in the literature 
(and will be reviewed below),
those for going from the theory to the band-power are not.  
The point of this paper is to provide those directions --- directions which
are clearly essential to the confrontation.

We now explicitly define the band-power method \cite{bondbp} which has
been used by many authors, {\it e.g.}, \cite{BJKII,Bartlett,bppapers}.  
In the simplest case of a dataset that produces one band-power,
its calculation is conceptually straightforward:
the power spectrum is assumed to be flat ($\C_l$ independent of $l$,
where $\C_l \equiv {l(l+1)\over 2\pi}C_l$)
and the band-power estimate is taken to be the amplitude 
of $\C_l$ estimated from the data ({\it e.g.}, via likelihood analysis).
For dataset $B$,
let us call this maximum likelihood value, $\C_B$.  Let us further
assume that the uncertainty in $\C_B$ is Gaussian-distributed with
variance $\sigma_B^2$.  

Because theoretical power spectra are not flat, the relation between
a theoretical power spectrum and the prediction of that theory
for the $\C_B$ derived from data is non-trivial.  The theoretical
prediction is simply the expectation value of $\C_B$, given
that the theory is true.  
Since $\C_B$ is a determination of the amplitude of the power spectrum 
we will assume a linear dependence of its expectation value on
the power spectrum of the theory, specified by the band-power
window function, $W^B_l$:
\be
\langle \C_B \rangle = \sum_l \left(W^B_l/l\right)\C_l(a_p)
\ee
where $a_p$ are the parameters of the theory.  These parameters,
$a_p$, could be cosmological parameters ({\it e.g.}, $\Omega_b$,
$\Omega_{\Lambda}$, $H_0$, etc.) or parameters from a phenomenological
power spectrum.  Throughout we will assume that $W_l^B$ is normalized
so that $\langle \C_B \rangle = \C_l$ for $\C_l$ independent of $l$.

With the assumptions of independence and Gaussianity for the
uncertainty in $\C_B$ and the specification of the linear
relationship between $\C_l$ and $\langle \C_B \rangle$, 
it follows that the likelihood
of the parameters is maximized by minimizing the following $\chi^2$:
\be
\label{eqn:simplechisq}
\chi^2 = \sum_B (\sum_l \left(W^B_l/l\right) \c_l(a_p) - \C_B)^2/\sigma_B^2,
\label{eq:chisq}
\ee
This $\chi^2$ represents the confrontation between theory and data
that occurs at the meeting point of the band-power.
Use of this equation, or ones similar to it, is very efficient
and is what has been used in analyses of the constraints
placed on parameters due to available CMB data.  Note that 
$W^B_l$ projects the theory into the ``plane'' of the 
experiment --- or rather the same plane into which the experimental
data have also been reduced.

Previous work has focused on generalizing Eq.~\ref{eqn:simplechisq}
to take into account the non-Gaussianity and dependence
of the uncertainties in $\C_B$ \cite{BJKII,Bartlett}.  
Here we focus on the choice of $W^B_l$.  
It is often assumed to be equal to the
variance window function, $W_l^V$,
which is actually the diagonal element
of a window function {\it matrix} which specifies the relationship 
between the angular power spectrum and the covariance 
matrix of the signal, $S$:
\be
\label{eqn:sigmat}
S_{pp'} \equiv \langle s_p s_{p'} \rangle = \sum_l \C_l W_{l,pp'}/l.
\ee 
where $s_p$ is the signal contribution to the $p$th element of
a dataset and the brackets indicate ensemble average.  For
a single demodulation, all the
diagonal elements of the window function matrix are equal and that
is why we can speak of {\it the} diagonal element.  

While using the variance window function to calculate the signal covariance
matrix is correct (this is what the variance window function is defined to do,
see, {\it e.g.}, \cite{leshouches,whitesred,OSA}),
using it in Eq.~\ref{eqn:simplechisq}
is not (except in the special case specified below).  

It is perhaps worth emphasizing the prevalent use of the variance
window functions in equations
like Eq. 2.  All of the references in ref. [1,6] use it, as do all published
analyses of large numbers of band powers.  This use of $W_l^V$
is due to the fact that a large number of reports of band-power
constraints do not include $W_l^B$, but only $W_l^V$ (e.g. 
\cite{experiments}).

\section{The band-power window function}

A minimum-variance, unbiased, estimate of the power spectrum
is given by \cite{BJKI,teg}
\be
\label{eq:estimate}
\C_l = {1\over 2}F_{ll'}^{-1}{\rm Tr}\left[\left(\Delta \Delta -N\right)\left(S+N\right)^{-1}
{\partial S \over \partial \C_{l'}} \left(S+N\right)^{-1}\right]
\ee
where
\be
\label{eqn:Fisher}
F_{ll'}= {\rm Tr}\left[(S+N)^{-1} {\partial S \over \partial \C_l}
(S+N)^{-1} {\partial S \over \partial \C_{l'}}\right]
\ee
is called the Fisher matrix.  

If we are only interested in estimating the amplitude of a power
spectrum that we assume to be flat, ($\C_l = \C_B = $ constant)
we can rewrite the minimum-variance estimator for $\C_l$ 
(Eq.~\ref{eq:estimate}) as
\be
\label{eqn:data2bp}
\C_B={1\over 2}F_{BB}^{-1}{\rm Tr}\left[\left(\Delta \Delta - N\right)
\left(S+N\right)^{-1}
\sum_{l'}{\partial S \over \partial \C_{l'}} \left(S+N\right)^{-1}\right]
\ee
where $F_{BB} = \sum_{ll'}F_{ll'}$ (because $\partial S/\partial \C_B
= \sum_{l} \partial S/\partial \C_l$).  Equation ~\ref{eqn:data2bp}
can be viewed as the directions that take one from the data, $\Delta$, 
to the meeting point of the band-power.

We now must provide the directions to go from a theory to the
band-power.  
With the usual assumptions of Gaussianity and statistical isotropy,
theories are completely specified by their angular power spectrum, $\C_l$.
We are therefore after the expectation value of the $\C_B$ of 
Eq.~\ref{eqn:data2bp}, under the assumption that the true power spectrum
is $\C_l$.  
Calculation of the dependence of this expectation value on $\C_l$ will provide 
the directions we need.

The expectation value is easily calculated after noting that
$\langle \left(\Delta \Delta -N \rangle\right) = S = \sum_l\C_l
\left({\partial S \over \partial \C_l}\right)$; 
it is given by
\be
\langle \C_B \rangle = {1\over 2}F_{BB}^{-1}{\rm Tr}
\left[\sum_{ll'}\C_l{\partial S \over \partial \C_l} \left(S+N\right)^{-1}
{\partial S \over \partial \C_{l'}} \left(S+N\right)^{-1}\right]
\ee
which further simplifies to 
\be
\label{eqn:expect}
\langle \C_B \rangle = {\sum_{ll'} \C_l F_{ll'} 
\over \sum_{ll'}F_{ll'}} = \sum_l \left(W^B_l/l\right) \C_l
\ee
which implicitly defines the band-power window function:
\be
\label{eqn:filter}
W^B_l/l = {\sum_{l'} F_{ll'} \over \sum_{ll'}F_{ll'}}.
\ee

We have found our linear relationship between the expected
value of $\C_B$ and the assumed power spectrum, $\C_l$.
We note that it has a form we might have guessed --- an inverse-variance
weighted sum of the $\C_l$.  We can identify it as such 
because the Fisher matrix also serves as an approximation to
the inverse of the covariance matrix of the uncertainty in 
the $\C_l$ estimates.

For band-powers derived from an estimator, the derivation of the
band-power window function is quite straightforward: one simply
calculates the expectation value, given $\C_l$, as done above.
However, band-powers are often determined instead by finding the
maximum of the likelihood function, rather than by the quadratic
estimator of Eq.~\ref{eq:estimate}.  The maximum-likelihood estimate
is a complicated, non-quadratic, function of the data and its
expectation value is not easy to calculate.  In fact, for a
maximum-likelihood estimate the band-power window function is
ill-defined because the relationship between the
maximum-likelihood and $\C_l$ is non-linear.

Nevertheless, the above expressions for the band-power window function
are still useful for maximum-likelihood band-power estimates.  This is
due to the relationship between likelihood analysis and the quadratic
estimator of Eq.~\ref{eq:estimate} pointed out in \cite{BJKI}: used
iteratively, Eq.~\ref{eq:estimate} results in the band-power that
maximizes the likeliihood.  If the $\C_l$ assumed for the right-hand
side of Eq.~\ref{eq:estimate} is at least roughly consistent with the
data (and ``smooth'' \cite{BJKII,BJKI,uros,OSH}) then a single iteration
of Eq.~\ref{eq:estimate} will produce a very good approximation to the
maximum-likelihood.  Thus, the band-power window function is
appropriate for $\C_l$ sufficiently close to the most likely power
spectrum.  Further away, non-linear corrections will become important.
One could, in principle, calculate these non-linear corrections, but
the improved precision is probably not worth the additional complication.

We emphasize that Eq.~\ref{eqn:expect} truly does specify a linear
relationship between the quadratically estimated, $\langle \C_B
\rangle$, and $\C_l$.  One might suspect that there are other
dependencies on $\C_l$ hidden in the window function itself.  However,
the Fisher matrix that appears twice in Eq.~\ref{eqn:filter} is that
for a flat power spectrum with amplitude $\C_B$, and does not depend
on $\C_l$.

\section{Three Examples}

Consider an experiment that maps the whole sky with
a Gaussian beam with full width at half max $= \sqrt{8\ln 2 }\sigma_b$ 
and a uniform noise
level specified by a weight-per-solid angle, $w$.
In this case of uniform noise and full-sky coverage, the
Fisher matrix can be calculated analytically and is given
by \cite{mythesis}
\be
\label{eqn:anafilt}
F_{ll'} = {2l+1 \over 2} \left[\C_l + {l(l+1) \over 2\pi w B^2(l)}\right]^{-2}
\delta_{ll'}
\ee
where $B(l) = e^{-l^2\sigma_b^2/2}$.  Due to the $\delta_{ll'}$, the
sum over $l'$ is trivial and the band-power window function is
\be
W^B_l/l \propto {2l+1 \over 2} \left[\C_l + {l(l+1) \over 2\pi w B^2(l)}\right]^{-2}.
\ee
Thus, for $\C_l$ constant, we see that the band-power window 
function for this map is proportional to $l^2$ at low $l$
and then eventually drops very rapidly at higher $l$ where it is
proportional to $B^{4}(l)/l^2$.

This behavior of $W^B_l$ is intuitively reasonable.  Cosmic
variance is the reason that the very low $l$s are less important to
the overall determination of the band-power, and instrument
noise suppresses the importance of the very high $l$s.

Contrast this behavior to that of the variance window function.  For
a map, $W^V_l$ is simply given by the square of 
the spherical harmonic transform of the beam, 
$B(l)$.  Therefore
\be
W^V_l \propto B^2(l).
\ee
Note that this implies that the
most important moments are the ones at lowest $l$!  
Further, there is no dependence on the noise level.  For
the band-power window function we see that as the noise is 
lowered ($w$ raised), the importance of the higher $l$ moments increases.

Our second example is for a dataset with $n$ points that have
no signal or noise correlations.  We leave it as an
exercise for the reader to show that in this case, 
$W^B_l = W^V_l$.  That is, in the absence of correlations, the
two window functions are equivalent.

\begin{figure}[bthp]
\plotone{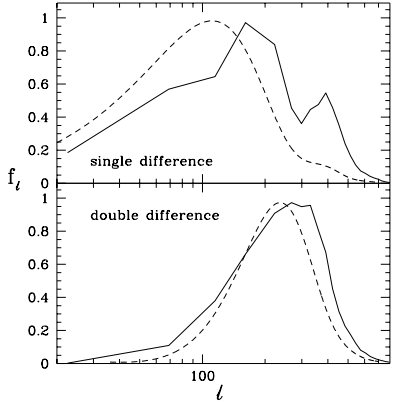}
\caption[Window functions]{\baselineskip=10pt \small The band-power 
window functions (solid lines) and variance window
functions (dashed lines) for the single and double difference
MSAM beam maps ($f_l \equiv W_l/l$).  The normalization here is arbitrary. 
}
\label{fig}
\end{figure}

Our final example is from the Medium Scale Anisotropy Measurement
(MSAM) 3-year dataset \cite{companion}.  This dataset was reduced to
measurements of the sky with two different beam maps, called
single-difference and double-difference.  The high signal-to-noise and
dense sampling of the dataset mean that it is sensitive to $\C_l$ at
somewhat higher values of $l$ then one would infer from the variance
window function.  See Fig. ~\ref{fig}.  For the single-difference
measurements, the results are especially striking: the peak is shifted
from $l=120$ to $l=160$ and at $l=400$, where there is a second local
peak, $W^B_l$ is about 5 times larger than $W^V_l$.  The band-power
window functions were calculated assuming a flat power spectrum with
amplitude consistent with the data of $\C_B = 2000 (\muk)^2$.

\section{Multiple Bands}

In the above we have assumed a flat power spectrum.
This is overly restrictive given that we wish to determine
the presence of features in this power spectrum!  However,
the above easily generalizes to the case where the power
spectrum is determined in multiple flat bands, each of
finite extent in $l$.  

Some datasets have sufficient dynamic range to estimate the
power spectrum in more than one band.  For these datasets we
can parameterize the power spectrum, $\c_l$,  with the
power in bands enumerated by the subscript $B$:
\be
\label{eq:cl}
\c_l = \sum_B \chi_{_{B(l)}} \C_B.
\ee
Where $\C_B$ is the amplitude of the power spectrum within band $B$ 
and
\be
\chi_{_{B(l)}}  = \left\{ 
\begin{array}{r@{\quad:\quad}l}
1 & l_<(B) \ < \ l \ < \ l_>(B) \\
0 & {\rm otherwise}
\end{array} 
\right\}
\ee
where $l_<(B)$ and $l_>(B)$ delimit the range of band $B$.
For each of these bands we can
calculate a band-power window function via
\be
W^B_l/l = {\sum_{l'} F_{ll'} \over \sum_{ll'} F_{ll'} }
\ee
where the sums over $l'$ run only from $l_<(B)$ to $l_>(B)$.

We could remove the need for window functions by making the bands very
narrow since sufficiently narrow bands ensure that the sensitivity
to each $\C_l$ within the band is approximately independent of $l$.
However, making the bands too narrow makes the error bars
very large and highly correlated.  This is undesirable for two
reasons.  First, it hinders visual interpretation of the results
and second, the larger the error bars, the more important are the 
non-Gaussian aspects of the distribution.  And although the ansatzes
of \cite{BJKII} for this non-Gaussian distribution have been shown to work
quite well in some cases, it is not clear how well they work in all
cases.  Therefore, broad bands may be desirable and the sensitivity
to $\C_l$ may vary appreciably across the band.  
In such cases window functions, $W^B_l$, tell us the in-band sensitivity. 

For experiments with small sky coverage, calculation of the elements
of the Fisher matrix at every $\ell$ is not necessary.  If the
largest extent of the field is $\Delta \theta$ then $P_l(\cos \theta)$
and $P_{l+\delta l}(\cos \theta)$ are close to indistinguishable
if $\delta l \la \pi/\Delta \theta$.  Thus one can choose
a fine binning, enumerated by $b$, within each coarse band, $B$ and
assume that $F_{ll'}$ is constant for all $l$ and $l'$ within band $b$.

It some times may not be practical to calculate the Fisher matrix
for individual multipole moments.  It may be easier
to parameterize the power
spectrum in terms of these fine bins,
\be
\label{eq:clfine}
\c_l = \sum_b \chi_{_{b(l)}} \C_b
\ee
and then calculate $F_{bb'}$ where
\be
\label{eqn:finefishmat}
F_{bb'}=
{\rm Tr} \left[(S+N)^{-1} {\partial S \over \partial \C_b}
(S+N)^{-1} {\partial S \over \partial \C_{b'}}\right].
\ee
Then one assumes that $F_{ll'} = F_{bb'}/(\delta l(b) \delta l(b'))$,
where $\delta l(b)$ is the width in $l$ of fine band, $b$.
We divide by the widths because
${\partial S \over \partial \C_{l}} \simeq  {\partial S \over \partial \C_{b}}/\delta l(b)$.

\section{Other shapes}

It may be the case that an experiment reports a single measure
of the power, but does so assuming a non-flat power spectrum shape.
An historical example is the Gaussian
auto-correlation function.  Another possibility is that
of an experiment measuring near the damping tail of the power
spectrum, where assuming a flat shape may be a very bad
approximation.  Therefore we ask, if the amplitude of
a non-flat power spectrum is estimated from the data, how
can we calculate the theoretical predictions for this quantity?

To frame the question more precisely, we assume that $Q$ is
calculated from a data set via
\be
\label{eqn:quadestQ}
Q = {1\over 2}F_{QQ}^{-1}{\rm Tr}
\left[\left(\Delta \Delta - N\right)\left(S+N\right)^{-1}
\sum_{l'}{\partial S \over \partial Q} \left(S+N\right)^{-1}\right]
\ee
where the power spectrum is assumed to be of the form 
$Q \C_l^{\rm shape}$.  Using similar manipulations as before,
we find that the expectation value of $Q$, under the assumption
that the true power spectrum is $\C_l$, to be
\be
\langle Q \rangle = {\sum_l \C_l \sum_{l'} F_{ll'}\C_{l'}^{\rm shape} \over
\sum_{ll'} F_{ll'}\C_{l}^{\rm shape} \C_{l'}^{\rm shape} } = 
\sum_l \left(W^Q_l/l\right)
\C_l.
\ee
Thus we have the prescription for comparing the estimated amplitude
to the predicted amplitude.  

Note that there is no clearly preferable means of converting the
estimate of the amplitude, $Q$, into a measure of average power.
One possible prescription is:
\be
\C_Q = Q \sum_l \left(W^Q_l/l\right) \C_l^{\rm shape} / \sum_l 
\left(W^Q_l/l\right).
\ee
Such a conversion is only useful for plotting purposes.  The ambiguity
in the choice of normalization does not disturb our ability
to confront data with theory.
As long as we know $W^Q_l$, and the shape assumed, 
$\C_l^{\rm shape}$, we can make the theoretical prediction for $Q$.

\section{Data Reporting Recipe}

To summarize, our prescription for reporting power spectrum
constraints is as follows:\\
1)  Parameterize the power spectrum via Eq.~\ref{eq:cl} for some
choice of bands.\\
2)  Find the $\C_B$ that maximize the likelihood function.\\
3)  Calculate the curvature matrix for these bands, ${\cal F}_{BB'}$,
and also the log-normal offsets $x_B$ defined in \cite{BJKII}.\\
4)  Calculate the band-power window functions, $W^B_l$, from 
$F_{ll'}$ (which can be calculated either via Eq.~\ref{eqn:Fisher} 
or Eq.~\ref{eqn:finefishmat}).

Steps 1 through 3 have been spelled out in more detail in
\cite{BJKII}.  We have no general prescription for the best
parameterization of the power spectrum to use for a given dataset
({\it e.g.}, how many bands to use and whether or not to assume a flat
shape).  We expect that assuming a flat spectrum across each band will
be a reasonable choice in most situations.  Whatever parameterization
is chosen, the analyst should ensure that in addition to the estimate
and its uncertainties, he or she also provide the means with which to
convert a theoretical power spectrum into a prediction for that
estimate.

\section{Discussion}

As emphasized in \cite{BJKII}, approximate methods for simultaneous
analysis of all relevant CMB data are a practical necessity.  The use
of band-power window functions, instead of variance window functions
will improve the validity of the commonly used method of radical
compression to band-powers.  

The persistence of the use of variance window functions as opposed to
band-power window functions (without even acknowledgment that this is,
at best, an approximation) is possibly attributable to the fact that
maximum-likelihood estimates have a very complicated dependence on the
data and, in fact, do not even have strictly well-defined band-power
window functions.  This conjecture is supported by the fact that
analyses using quadratic estimators have not suffered from this
confusion, while almost all of those using likelihood analysis
have.  It should also be noted that in analyses of galaxy redshift
surveys, where quadratic estimators are generally used to estimate the
matter power spectrum, $P(k)$, the correct form of the window function
is generally used, {\it e.g.} \cite{hamiltonb}.

As signal-to-noise rises, it becomes increasingly important to use
the correct window function.  This is because the power spectrum estimate
becomes sensitive to more pairs of data points than just the diagonal
ones---even the off-diagonal ones with very small signal matrix elements.
One can see from Eq. (7) that in the limit that $N \rightarrow 0$, all
pairs (normalized to their expected signal) get weighted equally.  In
this limit the much more numerous off-diagonal pairs are extremely
important to the determination of the band-power.  As a rough guide,
one can compare the size of the largest off-diagonal terms to the
noise level to determine how well the variance window function will
approximate the band-power window function.  Also note that the
approximation is generally worse for map datasets than difference
datasets due to the fact that differencing reduces the off-diagonal
correlations.

Steps toward the proper definition of the band-power window function
were taken in \cite{OSA} where the diagonal elements of the window
function matrix in the s/n basis were used as a means of determining
the sensitivity of an experiment to the power spectrum.  Working in
the signal-to-noise eigenmode basis \cite{bondbp,bunn} reduces the
correlations, and we have seen that in the absence of correlations,
the band-power window function is the variance window function.  A similar
procedure was used to calculate the window functions for the band-powers
determined from the QMAP maps \cite{Qmap}.

As individual datasets become more powerful in their ability to
determine $\C_l$, they will report constraints in very narrow
bands, decreasing the need for window functions which describe the in-band
sensitivity.  However, a useful role for window functions may remain for quite
some time at both the low $l$ and high $l$ extremes of a datasets'
power spectrum sensitivity.  At these extremes one will need broad
bands in order to have small error bars, and thus one will wish to
know the shape of the in-band sensitivity.

\acknowledgments
I am grateful to J.~R. Bond for useful conversations on this
subject, to G. Wilson for supplying the $\partial S / \partial \C_b$ 
necessary to make the MSAM band-power window functions in Fig.~\ref{fig}, 
and to G. Wilson and an anonymous referee who both helped to clarify
the text.  I am supported by the DOE, NASA grant NAG5-7986 and NSF
grant OPP-8920223.

\end{document}